\documentclass[aps,prl,twocolumn,amsmath,amssymb,superscriptaddress]{revtex4-2}
\usepackage{graphicx}
\usepackage{amsmath}
\usepackage{amssymb}
\usepackage{appendix}

\def\be{\begin{equation}}
\def\ee{\end{equation}}
\def\ba{\begin{eqnarray}}
\def\ea{\end{eqnarray}}
\def\go{\mathrel{\raise.3ex\hbox{$>$}\mkern-14mu
             \lower0.6ex\hbox{$\sim$}}}
\def\lo{\mathrel{\raise.3ex\hbox{$<$}\mkern-14mu
             \lower0.6ex\hbox{$\sim$}}}

\def\cR{{\cal R}}

\begin{document}

\title{Magnetic Power in Binaries: From Coalescing Binary Neutron
  Stars to Close-in Exoplanetary Systems}

\author{Dong Lai}
\affiliation{Tsung-Dao Lee Institute, Shanghai Jiao Tong University, Shanghai 201210, China}
\affiliation{Department of Astronomy, Cornell Center for Astrophysics
  and Planetary Science, Cornell University, Ithaca, NY 14853, USA}

\author{Tu Guo}
\affiliation{Tsung-Dao Lee Institute, Shanghai Jiao Tong University, Shanghai 201210, China}

\author{Nicholas Corso}
\affiliation{Department of Astronomy, Cornell Center for Astrophysics
  and Planetary Science, Cornell University, Ithaca, NY 14853, USA}

\begin{abstract}
A binary system containing two magnetic components (``stars'') can be
modeled as a circuit powered by the orbital motion, with the magnetic
field lines linking the two stars serving as wires. When the total
resistance in the binary components is small -- as found in many
astrophysical systems such as binary neutron stars and extra-solar
hot-Jupiter or super-Earth systems -- the connected field lines are
twisted and open up in a quasi-cyclic manner, leading to
quasi-periodic energy dissipation and flares. We derive analytic
expressions for the average and episodic magnetic powers from such
systems, providing a quantitative description on how the powers depend
on the magnetic orientations of the two stars. The power can be much
larger than in the case where one of the stars has a negligible
magnetic field. It may manifest as pre-merger quasi-cyclic flares in
coalescing neutron-star binaries and as orbit-regulated radio
emissions in exoplanetary systems.
\end{abstract}

\maketitle

%\section{Introduction}
{\bf Introduction}.
The electrodynamic coupling of a magnetized body to its companion has
a long history. Goldreich and Lynden-Bell \cite{Go69} developed a DC
circuit model for the magnetic interaction between Jupiter and its
satellite Io. In this model, Io's orbital motion through Jupiter's
rotating magnetosphere generates an electromotive force, driving a DC
current between the two bodies, with the flux tube linking them acting
as the conducting wire. This model helped explain Jupiter's decametric
radio emission that is modulated by Io's orbital motion (e.g., \cite{Cl96}).

The Jupiter–Io system represents a high-resistance binary pair, where
the flux tube connecting the planet and satellite is steady, with
dissipation balancing the flux twisting driven by spin and orbital
motion. By contrast, in most binary systems — including star–star and
star–planet systems (e.g.,
\cite{Lanza09,Laine12,Strugarek14,Strugarek16,Strugarek17}) and binary
neutron stars or white dwarfs
\cite{Hansen01,Lyutikov19,Most20,Most22,Most23,Mahlmann25} — the resistance associated
with the binary components is likely small. Dissipation then occurs
in the tenuous plasma between and around the binary, in the form of
reconnection-driven flares \cite{Most20,Beloborodov21}, and the
circuit connecting the two bodies is expected to be in a quasi-cyclic
state \cite{Lai12,Most22,Most23}.

The Jupiter–Io system can be modeled as a magnetic body (planet)
interacting with an unmagnetized conductor (Io). Many binary systems,
however, have non-negligible magnetic fields in both components, so
that the linked flux — and hence the magnetic power — depends on the
relative orientation of the two magnetic dipoles. What, then, is the magnetic
power in such systems?

In recent years, numerical simulations of such electrodynamic
interactions have been carried out in different contexts, from
close-in star–planet systems \cite{Strugarek16} to magnetic
neutron-star binaries \cite{Most20,Most22,Most23}. Although these
simulations address specific magnetic configurations and parameters,
they support the general picture of a quasi-cyclic DC circuit
established by the field lines linking the two components, with the
linkage interrupted by magnetic dissipation. However, a quantitative,
description of the power for generic magnetic comfigurations and
parameters is still lacking.

Lai \cite{Lai12} derived the maximum magnetic power for a "magnetic +
non-magnetic" system in the low-resistance limit. Here we generalize
the analysis to binary systems in which both components are magnetic,
and obtain the average and episodic powers as functions of the
relative dipole orientation.

%%%%%%%%%%%%%%%%%%%%%%%%%%%%%%%%%%%%%%%%%%%%%
%\section{Magnetic Torque and Power}
{\bf Magnetic Torque and Power}.
Consider a binary system consisting of two stars: the primary
has mass $M_1$, radius $R_1$, and magnetic
dipole moment $\mu_1$, and the secondary has $M_2$, $R_2$ and
$\mu_2$. We take $\mu_1\ge \mu_2$. The orbital separation is $a$, and the
orbital angular frequency is $\Omega$. The polar magnetic field strength at
surface of the primary is $B_1=2\mu_1/R_1^3$, and similarly $B_2=2\mu_2/R_2^3$.
The binary system is embedded in a tenuous plasma.
For simplicity, we neglect stellar spin (see \cite{Cherkis21}).

%%%%%%%%%%%%%%%%%%%%%%%%%
\begin{figure*}[t!]
\centering
\vskip -3.5truecm
\includegraphics[width=1.03\textwidth]{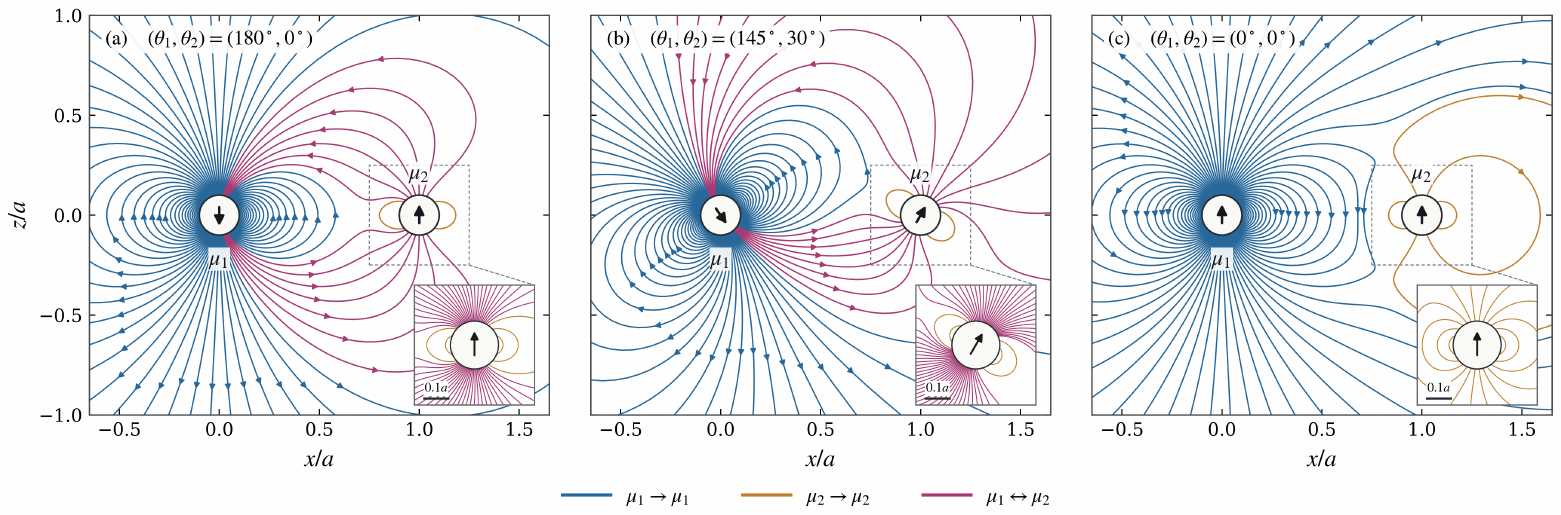}
\vskip -4.0truecm
\caption{Magnetic field configurations of two dipoles in vacuum. The two
  dipoles ($\mu_1$ and $\mu_2$) are separated by a distance $a$ (the
  binary separation), are oriented from $z$-axis (the binary orbital
  angular momentum axis) by the angles $\theta_1$ and $\theta_2$
  (measured clockwise from the $z$-axis).
  In these examples, the two dipole vectors are in the same plane, and
  $\mu_2/\mu_1=0.01$ is assumed. In general, there are three types of
  field lines: closed field lines of $\mu_1$ (blue), closed field lines
  of $\mu_2$ (orange), and linked field lines that connect the two
  dipoles (purple). For the aligned configuration (the right panel),
  there are no directly linked field lines in the absence of
  dissipation.}
    \label{fig1}
\end{figure*}
%%%%%%%%%%%%%%%%%%%%%%%%

The interaction between the two stars can be understood as
follows. Suppose there are magnetic field lines connecting the two
stars.  An effective DC circuit is set up, with the EMF driven by the
orbital motion, and the linked field lines serving as wires
\cite{Go69,Lai12}.  When the system's total resistance satisfies
$\cR_{\rm tot} \lo 16v_{\rm rel}/c^2$, where $v_{\rm rel}$ is the
relative orbital velocity, a steady-state DC circuit can not be
maintained \cite{Lai12}.  Based on the studies of solar flares and
magnetic star-disk systems, where magnetic flux tubes connecting
different regions of the system are twisted by differential motion
(e.g., \cite{aly85,aly90,van94,Lo95,Uz02}),
the magnetic binary system may exhibit the following quasi-cyclic behavior:
\begin{itemize}
\item A magnetic flux tube naturally connects the two stars,
  establishing magnetic linkage between them.
\item The linked fields are twisted by the orbital motion, generating
  a toroidal field from the linked poloidal field.
\item Once the toroidal magnetic field becomes comparable to the
  poloidal field, the fields inflate as magnetic bubbles and the flux
  tube breaks, disrupting the magnetic linkage.
\item Reconnection between the inflated field lines relaxes the twist
  and restores the magnetic linkage between the two stars.
\item The cycle restarts.
\end{itemize}
This cyclic behavior has been observed in recent 3D simulations of
specific magnetic binary systems \cite{Strugarek16,Most22,Most23}.

To estimate the magnetic power based on the above picture, we first
consider the case where both magnetic dipoles are
perpendicular to the orbital plane and are anti-aligned with each
other. A flux tube connects the two stars (see Fig.~\ref{fig1}).  Note that
because of the magnetic interaction, the effective size of the secondary,
$r_{m2}$, is determined by magnetic pressure balance (assuming that
the magnetic pressure dominates over the plasma ram pressure and
thermal pressure), i.e. $\mu_1/(a-r_{m2})^3 = \mu_2/r_{m2}^3$, giving
\be
{r_{m2}\over a}={\lambda\over 1+\lambda},\quad
{\rm with}\,\,\lambda\equiv(\mu_2/\mu_1)^{1/3}.
\ee
The region $r_2\lesssim r_{m2}$ (where $r_2$ is the radius measured from the
center of $M_2$) defines the secondary's magnetosphere.
Let $B_{z1}(a)=\mu_1/a^3$ be the vertical (perpendicular to the orbital plane)
magnetic field from the primary at the location of the secondary.
The azimuthal magnetic force per unit area (measured across the flux tube)
acting on the secondary is $B_{z1}B_{\phi +}/(2\pi)$, where $B_{\phi +}$ is the
toroidal field on the upper side the flux tube around the secondary.
For a given twist $\zeta_\phi\equiv -B_{\phi +}/B_{z1}$,
the magnetic torque on the secondary is
\be
T_{\rm mag}\sim -{1\over 2\pi}\zeta_\phi B_{z1}^2 \pi r_{m2}^2 a
\sim -{\zeta_\phi\over 2\pi} a B_{z1}\Phi_L,
\ee
where $\Phi_L\sim B_{z1}\pi r_{m2}^2$ is the linked magnetic flux.

In this picture, the mechanical energy of the binary is converted to the
magnetic energy (in the form of toroidal field) at the rate
$-T_{\rm mag}\Omega$. When the twist reaches $\zeta_\phi=\zeta_{\phi,{\rm max}}
\sim \pi$, the magnetic field inflates rapidly, generating a magnetic bubble (``flare'').
The time-averaged magnetic dissipation power is then
\be
{\bar L}_{\rm mag} =-{\bar T}_{\rm mag}\Omega
\sim {\bar\zeta_\phi\over 2\pi} a \Omega B_{z1}\Phi_L,
\ee
where the time-averaged twist $\bar\zeta_\phi\simeq \zeta_{\phi,{\rm max}}/2
\sim \pi/2$.

Thus, for general binary/magnetic geometry, we suggest the following parametrization
for the time-averaged magnetic power and torque:
\ba
&&{\bar L}_{\rm mag}=-{\bar T}_{\rm mag}\Omega\nonumber\\
&&\qquad\quad ={1\over 2}\,\bar\zeta_\phi f_\Phi\, a\Omega \left({\mu_1\over a^3}\right)^2 r_{m2}^2\nonumber\\
&&\qquad \quad={1\over 2}\,\bar\zeta_\phi f_\Phi \left({\lambda\over 1+\lambda}\right)^2
\Omega\,{\mu_1^2\over a^3},
\label{eq:Lmag}
\ea
where $f_\Phi$ is the dimensionless flux parameter
\be
f_\Phi\equiv {\Phi_L\over \pi r_{m2}^2 B_{z1}(a)}
={\Phi_L\over \pi r_{m2}^2 \mu_1/a^3}.
\label{eq:fphi}\ee

The magnetic energy is released in quasi-periodic episodic flares.
The stored energy in the twisted magnetic field is given by
\be
{\cal E}_{\rm twist}=\int_0^T dt\, L_{\rm mag}
\simeq {1\over 4}\,\zeta_{\phi,{\rm max}}^2\,f_\Phi \left({\lambda\over 1+\lambda}\right)^2
{\mu_1^2\over a^3},
\ee
where $T$ is the the twist-and-release period, and
we have used $\zeta_\phi\simeq\Omega t$ and $\zeta_{\phi,{\rm max}}\simeq \Omega T$.
If this energy is released on a short dissipative timescale $t_{\rm diss}$
($\ll 1/\Omega$), the flare luminosity is
\be
L_{\rm flare}\simeq {\cal E}_{\rm twist}/t_{\rm diss}.
\ee

%%%%%%%%%%%%%%%%%%%%
\begin{figure}
\vskip -4truecm  
\hspace*{-1.5cm}\includegraphics[angle=0,scale=0.54]{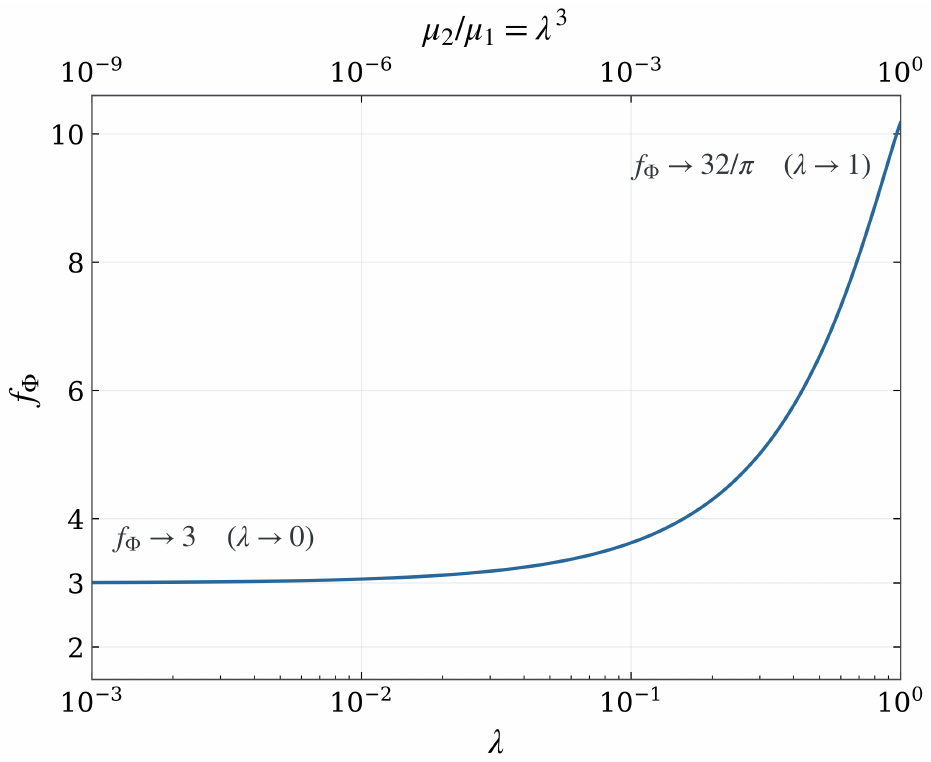}
\vskip -4.2truecm  
\caption{The dimensionless linked magnetic flux (defined by Eq.~\ref{eq:fphi})
  as a function of $\lambda=(\mu_2/\mu_1)^{1/3}$ for two anti-aligned dipoles
  ($\theta_1=180^\circ$ and $\theta_2=0$). See the left panel of Fig.~\ref{fig1}.}
\label{fig2}
\end{figure}
%%%%%%%%%%%%%%%%%%%

%%%%%%%%%%%%%%%%%%%%%
\begin{figure}
\vskip -3truecm  
\hspace*{-1.cm}\includegraphics[angle=0,scale=0.5]{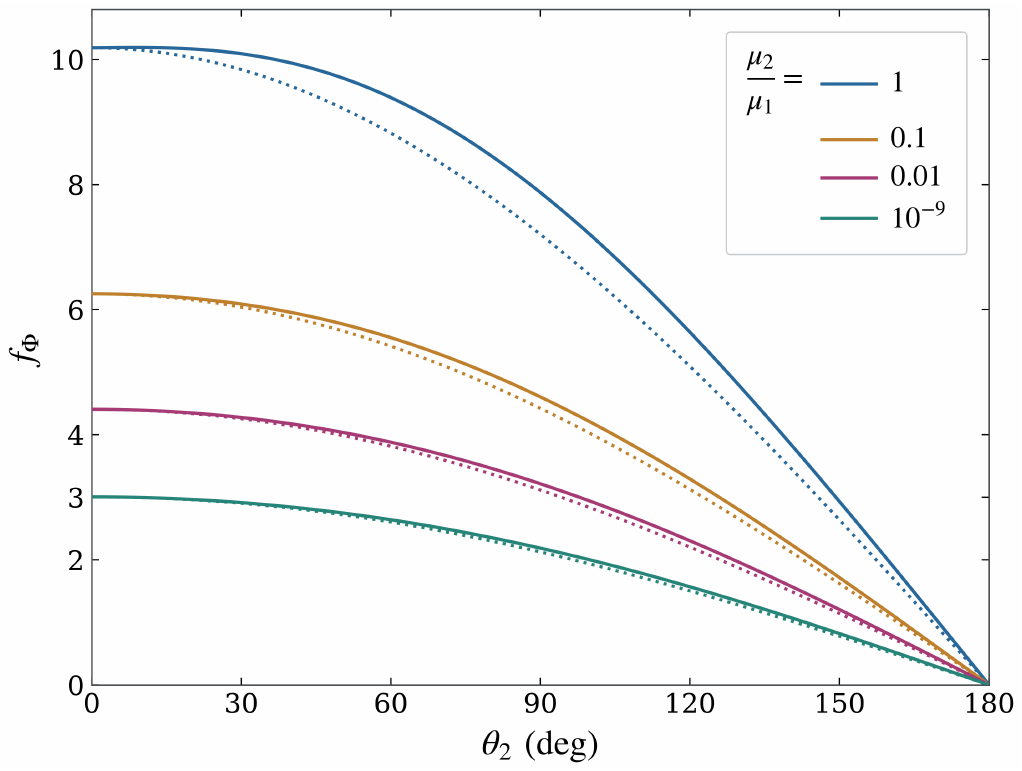}
\vskip -3.8truecm
\caption{The dimensionless linked magnetic flux (defined by Eq.~\ref{eq:fphi})
  of two dipoles as a function of $\theta_2$, for several values of
  $\mu_2/\mu_1$, assuming $\theta_1=180^\circ$. The dotted lines correspond
  to the function $f_\Phi(\theta_2=0)\cos(\theta_2/2)$.}
\label{fig3}
\end{figure}

\begin{figure}
\vskip -3truecm  
\hspace*{-1.cm}\includegraphics[angle=0,scale=0.5]{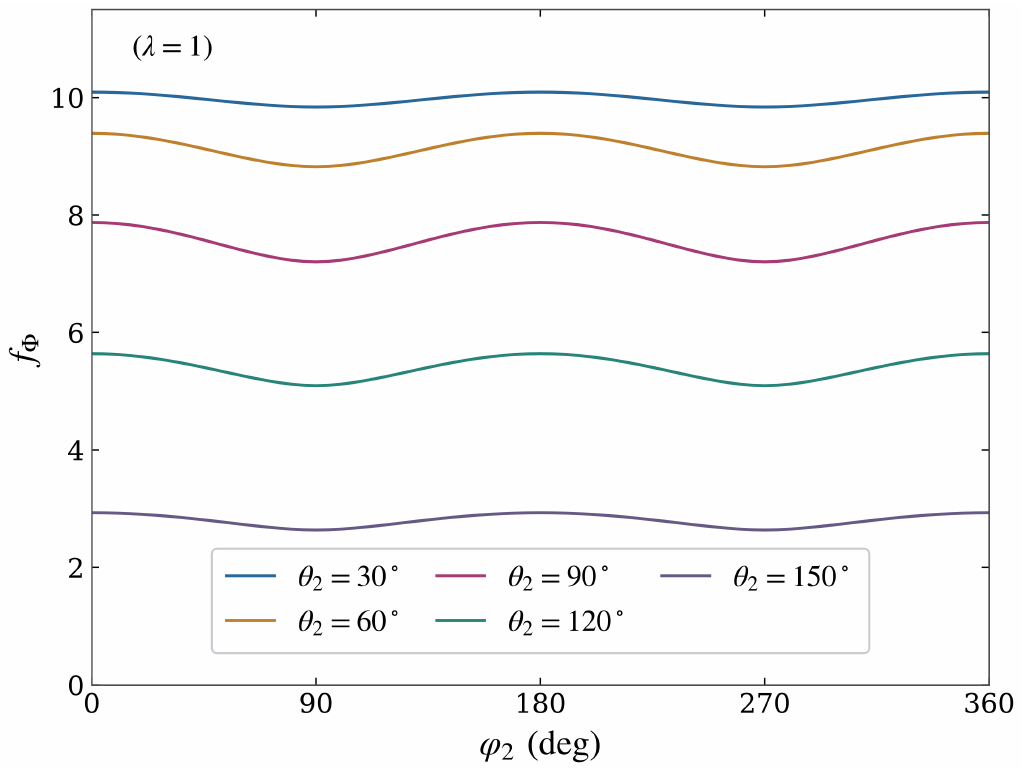}
\vskip -3.7truecm
\caption{The dimensionless linked magnetic flux (defined by Eq.~\ref{eq:fphi})
  of two dipoles as a function of $\varphi_2$ for several values of
  $\theta_2$ (assuming $\theta_1=180^\circ$ and $\lambda=1$).}
\label{fig4}
\end{figure}

{\bf Linked Magnetic Flux}.
Figure~\ref{fig2} shows $f_\Phi$ as a function of $\lambda$ in the range from
$\lambda \ll 1$ to $\lambda=1$ for the anti-aligned configuration.
We find that $f_\Phi$ increases from 3 for
$\lambda\rightarrow 0$ to $32/\pi$ for $\lambda=1$ (see Appendix).

In general, $f_\Phi$ depends on $\lambda$ and on the orientations of
the dipoles.  For concreteness, we consider $\theta_1=180^\circ$ (so
that $\boldsymbol{\mu}_1$ is anti-aligned with the orbital axis $\hat z$).
The orientation of $\boldsymbol{\mu}_2$ is specified by the polar angles $(\theta_2,
\varphi_2)$.
Figure~\ref{fig3} shows $f_\Phi$ as a function of $\theta_2$ (assuming $\varphi_2=0$)
for several values of $\mu_2/\mu_1=\lambda^3$. In the limit of $\lambda\rightarrow 0$,
$f_\Phi$ is given by (to within $5.8\%$) by (see Appendix)
\be
f_\Phi(\lambda,\theta_2)\simeq 3\cos(\theta_2/2),\quad (\theta_1=180^\circ,\,
\lambda\rightarrow 0).
\ee
For general values of $\lambda$, we see that $f_\Phi$ is still
well approximated (to within $10\%$) by
\be
f_\Phi(\lambda,\theta_2)\simeq f_\Phi(\lambda,\theta_2=0)\cos(\theta_2/2),\quad
(\theta_1=180^\circ),
\ee
with $f_\Phi(\lambda,\theta_2=0)$ ranges from $3$ to $32/\pi$ for $\lambda
\in [0,1]$. Note that for $\lambda\ll 1$, $f_\Phi$ is independent of
$\varphi_2$. Even for $\lambda=1$, the dependence of $f_\Phi$ on
$\varphi_2$ is quite weak (see Fig.~\ref{fig4}).

Clearly, maximum power is achieved for $\theta_2=0$ (anti-aligned configuration).
For $\theta_2=180^\circ$ (aligned configuration), there is no linked flux
in the ``natural'' (untwisted) state of the system. Also when
$r_{m2}\lesssim R_2$ (non-magnetic secondary star), there is no linked
flux in the absence of plasma dissipation. Such configurations tend to produce lower
level of magnetic power compared to systems with naturally linked flux.
For example, for system with $r_{m2}\ll R_2$, the
time-averaged magnetic power is expected to be \cite{Lai12}
\be
   {\bar L}_{\rm mag}^{(0)}\simeq {1\over 2}{\bar\zeta_\phi}
   \Omega\,{\mu_1^2 R_2^2\over a^5}.
\ee
For aligned system with magnetic components,
turbulent dissipation could arise from Kelvin-Helmholtz instability
at the interface of the stellar magnetospheres \cite{Mahlmann25} -- this effect is
no captured in our analysis above.

{\bf Effect of Finite Stellar Sizes.}  In the above we have assumed
$a\gg R_1,R_2$ and $r_{m2}\gtrsim R_2$.  Since magnetic interaction is
likely most important when $a$ is not much larger than $R_1$ and
$R_2$, it is necessary to consider the effect of finite stellar sizes.
For concreteness, consider the anti-aligned configuration
($\theta_1=180^\circ$ and $\theta_2=0$). Neglecting high-order (quadrupole or
higher) effects, the magnetic field of $\mu_1$
acting on the secondary (assumed to be a perfect conductor) induces a
dipole $\mu_{2,{\rm ind}}=\mu_1 (R_2^3/2a^3)$ in the direction aligned
with $\boldsymbol{\mu}_1$ (i.e. opposite to $\boldsymbol{\mu}_2$);
similarly, the primary attains an induced dipole
$\mu_{1,{\rm ind}}=\mu_2(R_1^3/2a^3)$.  Thus, in the natural (untwisted)
state, the poloidal field of the system is produced by two effective
dipoles
\ba &&\mu_{1,{\rm
    eff}}=\mu_1-(R_1^3/2a^3)\mu_2,\\ &&\mu_{2,{\rm
    eff}}=\mu_2-(R_2^3/2a^3)\mu_1.  \ea
With this, the calculation of the linked flux can proceed in the same way as above.

As an example, for systems with $\mu_1=\mu_2$ and $R_1=R_2$, we find
\be
{\bar L}_{\rm mag}\simeq {1\over 8}\,\bar\zeta_\phi f_\Phi 
\Omega\,{\mu_1^2\over a^3}\left(1-{R_1^3\over a^3}\right),\quad
(\lambda=1,~{\rm anti-aligned}).
\ee

%%%%%%%%%%%%%%%%%%%%%%%%%%%%%%%%%%%%%%%%%%%%%%%%%%%%%%%
%\section{Application: Neutron Star Binaries}
{\bf Application: Neutron-Star Binaries}.
Gravitational wave (GW) emission drives the orbital decay of the neutron star
(NS) binary, with timescale
\be
t_{\rm GW}=\left({a\over |\dot a|}\right)_{\rm GW}={5c^5a^4\over 64G^3M_1^3 q(1+q)}
\simeq 1.47\ \mathrm{s}\left(\frac{a}{100\ \mathrm{km}}\right)^4,
\label{eq:tgw}\ee
where in last equality we have adopted $M_1=1.4M_\odot$ and mass ratio
$q=M_2/M_1=1$. Note that the gravitational wave frequency is
%$f_{\rm GW}=\Omega/\pi\simeq 194\ \mathrm{Hz}
%\left(\frac{a}{100\ \mathrm{km}}\right)^{-3/2}.
$f_{\rm GW}=193\ (a/100\ \mathrm{km})^{-3/2}$~Hz.

The time-averaged magnetic energy dissipation rate is
\ba
&& {\bar L}_{\rm mag}\simeq 1.9\times 10^{42}\,
\bar\zeta_\phi f_\Phi \left({2\lambda\over 1+\lambda}\right)^2
\left(\frac{B_1}{10^{13}\,\mathrm{G}}\right)^2\nonumber\\
&&\qquad \times \left({R_1\over 10\,{\rm km}}\right)^6
\left(\frac{a}{100\,\mathrm{km}}\right)^{-9/2}
\ \mathrm{erg\,s^{-1}}.
\ea
Compared to the GW power, the ratio is 
\ba
&&{{\bar L}_{\rm mag}\over L_{\rm GW}}
\simeq 1.1\times 10^{-10}
\bar\zeta_\phi f_\Phi \left({2\lambda\over 1+\lambda}\right)^2
\left(\frac{B_1}{10^{13}\ \mathrm{G}}\right)^2\nonumber\\
&&\qquad \times \left({R_1\over 10\,{\rm km}}\right)^6
\left(\frac{a}{100\ \mathrm{km}}\right)^{1/2}.
\ea
The magnetic power leads to a phase shift to the binary inspiral waveform,
given by\cite{Lai12}
\be
\Delta\Phi_{\rm GW}\simeq -2\pi \left({{\bar L}_{\rm mag}\over L_{\rm GW}}\right)
\left({dN\over d\ln f}\right)_0,
\ee
where
\ba
&& \left(dN\over d\ln f_{\rm GW}\right)_0={2\over 3}f_{\rm GW}t_{\rm GW}
={5c^5(1+q)^{1/3}\over 96\pi q M_1^{5/3}(\pi Gf_{\rm GW})^{5/3}}\nonumber\\
&&\qquad \simeq  190 \left(\frac{a}{100\ \mathrm{km}}\right)^{5/2}
\simeq 574 \left(\frac{f_{\rm GW}}{100\ \mathrm{Hz}}\right)^{-5/3}
\ea
is the usual leading-order point-mass GW cycles.
We see that $|\Delta\Phi_{\rm GW}|\ll 1$ even for $B_1$ as large as $10^{15}$~G.

On the other hand, the dissipation of the twisted fields is likely due to
fast magnetic reconnection (with Alfven speed $\simeq c$), thus $t_{\rm diss}\sim a/c$, 
and the flare luminosity is
\ba
&& L_{\rm flare}\simeq 4.7\times 10^{42}
\zeta_{\phi,{\rm max}}^2 f_\Phi \left({2\lambda\over 1+\lambda}\right)^2
\left(\frac{B_1}{10^{13}\,\mathrm{G}}\right)^2\nonumber\\
&&\qquad \times \left({R_1\over 10\,{\rm km}}\right)^6
\left(\frac{a}{100\,\mathrm{km}}\right)^{-4}
\left({a/c\over t_{\rm diss}}\right)
\ \mathrm{erg\,s^{-1}}.
\ea
For $\lambda=1$ (which gives $f_\Phi=32/\pi$) and $R_1=14.76$~km
(the radius adopted by \cite{Most20}), we find
\ba
&&L_{\rm flare}\simeq 1.2\times 10^{45}
\left({\zeta_{\phi,{\rm max}}\over \pi}\right)^2
\left(\frac{B_1}{10^{12}\,\mathrm{G}}\right)^2\nonumber\\
&&\qquad \times \left(\frac{a}{45\,\mathrm{km}}\right)^{-4}
\left({a/c\over t_{\rm diss}}\right)
\mathrm{erg\,s^{-1}},
\ea
in good agreement with the numerical results of 
\cite{Most20,Most22}. Note that $L_{\rm flare}$ refers
to the total energy dissipation rate; only a small fraction ($\sim 10^{-4}$)
may go into radio emission.

%%%%%%%%%%%%%%%%%%%%%%%%%%%%%%%%%%%
%\section{Close-in Exoplanetary Systems}
{\bf Application: Close-in Exoplanetary Systems}.
For a planetary system with $M_1=M_\star$, $R_1=R_\star$, $M_2=m_p$, $R_2=R_p$,
assuming $\lambda\ll 1$ (thus $f_\Phi\simeq 3$),
we find that the magnetic torque induces orbital decay at the rate
\ba
&& {\dot a\over a}\simeq -{\bar\zeta_\phi f_\Phi \lambda^2 \mu_\star^2 \over
m_p a^3 (GM_\star a)^{\!1/2}}\nonumber\\
%&&\quad \simeq 7.8\times 10^{-15}\ \mathrm{yr}^{-1}\,
%\bar\zeta_\phi \left(\frac{M_\star}{M_\odot}\right)^{\!\!-1/2}
%\left(\frac{B_\star R_\star^3}{\mathrm{G}\ R_\odot^3}\right)^{\!\!4/3}
%\nonumber\\
%&&\qquad \times \!\left(\frac{B_p R_p^3}{\mathrm{G}\ R_J^3}\right)^{\!\!2/3}
%\!\!\left(\frac{m_p}{M_J}\right)^{\!-1}
%\!\left(\frac{a}{0.04\ \mathrm{AU}}\right)^{-7/2}.
%%%
&&\quad \simeq 2.5\times 10^{-15}\ \mathrm{yr}^{-1}\,
\bar\zeta_\phi 
\left(\frac{M_\star}{M_\odot}\right)^{\!\!-1/2}
\left(\frac{B_\star R_\star^3}{\mathrm{G}\ R_\odot^3}\right)^{\!\!4/3}
\nonumber\\
&&\qquad \times \!\left(\frac{B_p}{\mathrm{G}}\right)^{\!\!2/3}
\!\!\left(\frac{{\rm g/cm}^3}{\rho_p}\right)
\!\!\left(\frac{R_J}{R_p}\right)
\!\left(\frac{a}{0.04\ \mathrm{AU}}\right)^{-7/2},
\label{eq:adot}\ea
where $\rho_p=3m_p/(4\pi R_p^3)$ is the mean density of the planet.
The above expression assumes
\be
{r_{m2}\over R_p}\simeq \left({\mu_p\over \mu_\star}\right)^{\!\!1/3}
{a\over R_p}
=\left({B_p\over B_\star}\right)^{\!\!1/3}
{a\over R_\star} \gtrsim 1.
\ee
If $r_{m2}/R_p\ll 1$ (non-magnetic planet), 
the orbital decay rate is \cite{Lai12}
\be
{\dot a\over a}\simeq
-{\bar\zeta_\phi \mu_\star^2 R_p^2\over a^5 m_p(GM_\star a)^{\!1/2}}.
\label{eq:adot0}\ee

Using these expressions, we see that a hot 
Jupiter ($R_p\sim R_J$) or super-Earth ($R_p\sim 0.2R_J$) in a $P\sim 3$~d
orbit ($a\simeq 0.04$~AU) around a solar-type star (with a
typical dipole field $B_\star$ of a few Gauss) will suffer negligible
orbital decay due to the magnetic torque.  An ultra-short-period Earth-like
planet (with $a\sim 0.01$~AU, $R_p\sim 0.1R_J$, $B_p\sim 1$~G; see
\cite{LeeChiang17,PuLai19}) around a high magnetic star (e.g. $B_\star\sim 100$~G)
would have $|a/\dot a|$ less than $10^9$~years.
If super-Earths migrate to their close-in locations in the early T-Tauri phase of the star,
when the stellar magnetic field is stronger (a few kG), they may experience
appreciable orbital evolution due to magnetic interactions.
We note that Eqs.~(\ref{eq:adot}) and (\ref{eq:adot0}) give the maximum orbital
decay rate (for the assumed parameter values);
if the planet has a sufficiently large internal resistance, the decay rate
will be smaller \cite{Lai12}.

The maximum energy stored in the twisted magnetic field is (again for $\lambda\ll 1$)
\ba
&&{\cal E}_{\rm twist}\simeq {3\over 4}\,\zeta_{\phi,{\rm max}}^2
\lambda^2 \,{\mu_1^2\over a^3}\nonumber\\
&&\qquad \simeq
1.04\times 10^{28}\ \mathrm{erg}\
\left({\zeta_{\phi,{\rm max}}\over \pi}\right)^2
\left(\frac{B_{\star} R_{\star}^{3}}{\mathrm{G}\,R_{\odot}^{3}}\right)^{4/3}\nonumber\\
&&\qquad\times \left(\frac{B_{p} R_{p}^{3}}{\mathrm{G}\,R_{J}^{3}}\right)^{2/3}
\left(\frac{a}{0.04\,\mathrm{AU}}\right)^{-3}.
\label{eq:Etwist_normalized}
\ea
For the canonical parameter values adopted in the above, this is
smaller than the energy release of solar flares
($10^{29}-10^{32}$~ergs) and of superflares from solar-type stars
($\go 10^{33}$~ergs; see \cite{Mae12}). For larger values of $B_\star$
and/or $B_p$, or for small $a$, the stored energy can be much
larger. Some fraction of this energy can be released in the radio band
and becomes detectable.  Continued search for the signature of
star-planet interaction would be useful (e.g.,
%\cite{Sh08,Pi11,fa12,Mi12,Callingham24,Ilin25,Pineda26}).
\cite{Sh08,Pi11,Callingham24,Ilin25,Pineda26}).

%%%%%%%%%%%%%%%%%%%%%%%%%%%%%%%%%%%
{\bf Discussion}.
Our analytical study of magnetic power in binaries extends previous
work to systems in which both binary components (``stars'') have
intrinsic dipole magnetic fields. When the total resistance in the
binary components is small — as is the case in most astrophysical
systems — the magnetic field lines that connect the two stars are
twisted by the orbital motion and open up once the twist angle exceeds
$\sim \pi$, leading to quasi-periodic magnetic energy dissipation and
flares. For a given binary separation and given magnetic dipole
moments, the magnetic power from the binary is directly proportional
to the magnetic flux linking the two stars in the natural
(``untwisted'') state. This provides a quantitative description of how
the power depends on the dipole orientations of the two stars (see
Fig.~\ref{fig3}).

Applying our result to coalescing neutron-star binaries, we find that
although the magnetic power is enhanced compared to the case in which
one of the stars has a negligible magnetic field, it has a negligible
effect on the orbital decay rate (thus the GW phase) even for
magnetar-level magnetic fields. Our calculated pre-merger magnetic
energy dissipation rate agrees well with the results of full numerical
simulations \cite{Most20,Most22}, subject to the uncertainties in the
reconnection rate of the fully twisted magnetic fields. Similarly,
applying our general result to close-in hot-Jupiter or super-Earth
extrasolar systems, we find that for typical stellar and planetary
fields (a few Gauss), magnetic interaction does not lead to
appreciable orbital evolution, with the possible exception of the
early T Tauri phase, when the stellar dipole magnetic field is
stronger than $10^3$~G.  The energy dissipation induced by the orbital
motion of a magnetic planet around a strongly magnetized star can
become comparable to the energies released in observed stellar flares,
and orbit-correlated radio emission may be generated.

Our circuit model with linked field lines between the stars requires
that the slippage of the flux tube during the round-trip Alfv\'en travel
time along the flux tube ($t_A\sim 2a/v_A$, where $v_A$ is the Alfv\'en
speed of the poloidal field) be less than $r_{m2}$, i.e. $\Omega t_A a
\lesssim r_{m2}$, or $2(GM_{12}/ac^2)^{1/2}\lesssim
(v_A/c)(r_{m2}/a)$ (where $M_{12}=M_1+M_2$). When this condition is not satisfied,
the ``Alfv\'en wing'' model may be applied: the disturbance generated by the
secondary's orbital motion propagates along the
``external'' field line as Alfv\'en waves and radiates away \cite{Drell65,CorsoLai26}.
Taking account of the internal magnetic field of the secondary,
the Alfv\'en power is (Eq.~22 in Ref.~\cite{CorsoLai26b})
\be
L_{\rm Alfven}\simeq {\alpha v_{\rm rel}^2\over 2\pi v_A} B_{z1}\Phi_L,
\ee
with $\alpha\simeq 1.68$ from numerical simulations. Here $v_{\rm rel}$ is the velocity
of the secondary relative to the ``external'' medium (If the stellar-wind velocity
from the primary star is negligible, then $v_{\rm rel}=a\Omega$
is simply the orbital velocity), and $\Phi_L$ is the magnetic field flux linked
to the external field -- in a binary, this external field is produced by the primary.
Comparing with Eq.~(\ref{eq:Lmag}), we see that
\be
{L_{\rm Alfven}\over {\bar L}_{\rm mag}}\simeq {\alpha\, a\Omega\over \bar\zeta_\phi v_A}.
\ee
These equations apply in the sub-Alfvenic regime ($v_{\rm rel}\lesssim v_A$), where 
$L_{\rm Alfven}$ is always less than $\bar L_{\rm mag}$. Thus, the general expressions
presented in this paper represent the maximum magnetic interaction power in binaries.

\bigskip
%\begin{acknowledgments}
{\bf Acknowledgments}.  The analytical derivations presented in the
Appendix are obtained with the help of AI and checked by the authors.
%\end{acknowledgments}

%%%%%%%%%%%%%%%%%%%%%%
% Bibliography -- entries in order of first citation (PRL style)
%%%%%%%%%%%%%%%%%%%%%%

\vskip 2truecm

%%%%%%%%%%%%%%%%%%%
\appendix

\section{Appendix A: Connected Flux in the Weak-Dipole Limit}
\label{app:weak-dipole}

We consider two point magnetic dipoles located at the origin and at
\(a\hat{\mathbf x}\), with moments
\begin{equation}
\boldsymbol{\mu}_1=-\mu_1\hat{\mathbf z},
\qquad
\boldsymbol{\mu}_2=\mu_2\hat{\mathbf e},
\end{equation}
where \(\mu_i>0\) and
%\(\hat{\mathbf e}=(-\sin u,\cos u,0)\).
\(\hat{\mathbf e}=(\sin u\cos\varphi_2,\sin u\sin\varphi_2,\cos u)\),
with $u=\theta_2$.
For \(\lambda=(\mu_2/\mu_1)^{1/3}\ll1\), we can set $\varphi_2=0$, and
the field of dipole 1 is
approximately uniform across the connecting region near dipole 2.
Writing \(\mathbf r=r\hat{\mathbf n}\) relative to dipole 2, the local
field is
\begin{equation}
\mathbf B(\mathbf r)
=
B_0\hat{\mathbf z}
+
\frac{\mu_2}{r^3}
\left[
3(\hat{\mathbf e}\cdot\hat{\mathbf n})\hat{\mathbf n}
-\hat{\mathbf e}
\right],
\qquad
B_0=\frac{\mu_1}{a^3}.
\end{equation}

On a ``separating'' sphere of radius \(R\), the radial field is
\(\mathbf B\cdot\hat{\mathbf n}
=
\mathbf v\cdot\hat{\mathbf n}\), where
\begin{equation}
\mathbf v
=
B_0\hat{\mathbf z}
+
\frac{2\mu_2}{R^3}\hat{\mathbf e}.
\end{equation}
Requiring the field to be tangent to the separating circle
\(r=R\), i.e.,
\(\mathbf r\cdot\mathbf v=0\), gives
\be
2t^2-t\cos u-1=0,
\ee
with
\begin{equation}
t\equiv\frac{\mu_2}{B_0R^3}
=\frac{\cos u+\sqrt{\cos^2u+8}}{4}.
\end{equation}

The leading linked flux is the outward flux through the hemisphere
\(S_+=\{r=R,\hat{\mathbf n}\cdot\mathbf v>0\}\):
\begin{equation}
\Phi_L
\propto
\int_{S_+}\mathbf B\cdot\hat{\mathbf n}\,dS
=
\pi R^2|\mathbf v|
=
\frac{3\pi\mu_1\lambda^2}{a}h(u),
\end{equation}
with
\be
h(u)=
\frac{\sqrt{1+4t^2+4t\cos u}}{3t^{2/3}},
\end{equation}
where \(dS\) is the surface-area element and the value at
\(u=\pi\) is understood by continuity. Consequently, the
dimensionless connected flux is
\begin{equation}
f_\Phi   \equiv
\frac{\Phi}
{\pi[\lambda/(1+\lambda)]^2\mu_1/a}
\simeq
3(1+\lambda)^2h(u).
\end{equation}

Using \(\cos u=2t-t^{-1}\), the angular dependence factors exactly as
\begin{equation}
h(u)
=
\cos\frac{u}{2}\,\mathcal R(u),
\qquad
\mathcal R(u)
=
\sqrt{
\frac{2(2t+1)}
{3t^{1/3}(t+1)}
},
\end{equation}
with
\be
1\le\mathcal R(u)\le
\frac{2^{5/3}}{3}
=1.05827\ldots
\ee
Neglecting this small correction gives
\begin{equation}
  f_\Phi
\simeq
3(1+\lambda)^2\cos\frac{u}{2}
\longrightarrow
3\cos\frac{u}{2}
=3\cos\frac{\theta_2}{2}
\qquad
(\lambda\to0).
\end{equation}
Replacing \(h(u)\) by \(\cos(u/2)\) underestimates the uniform-field
result by at most \(5.8\%\) for \(u<\pi\); both vanish at
\(u=\pi\).

%%%%%%%%%%%%
\section{Appendix B: Connected Flux of Perpendicular Anti-Aligned Dipoles}

Consider two point dipoles with centered at 
$\mathbf r_1=\mathbf 0$ and $\mathbf r_2=a\hat{\mathbf x}$. Their
moments are $\boldsymbol\mu_1=-\mu_1\hat{\mathbf z}$ and
$\boldsymbol\mu_2=\mu_2\hat{\mathbf z}$.
%where $\mu_1,\mu_2>0$ denote the magnitudes.
Both moments are perpendicular to the separation axis
and opposite to one another.
With angles measured clockwise from $+z$ in the $xz$ plane,
this configuration corresponds to $(\theta_1,\theta_2)=(180^\circ,0^\circ)$.
%We parameterize the moment ratio by $\lambda=(\mu_2/\mu_1)^{1/3}$.

For a field point $\mathbf r=(x,y,z)$, let $\mathbf R_i=\mathbf
r-\mathbf r_i$ and $R_i=|\mathbf R_i|$. The magnetic scalar potential
and field are therefore
\be
\mathcal V(\mathbf r)=\sum_{i=1}^{2}\frac{\boldsymbol\mu_i\cdot\mathbf R_i}{R_i^3}
=z\left(\frac{\mu_2}{R_2^3}-\frac{\mu_1}{R_1^3}\right),
\qquad
\mathbf B=-\boldsymbol\nabla\mathcal V.
\ee
The scalar potential identifies a useful cross-section of the
connecting field lines. We can use the equipotential $\mathcal V=0$ to
calculate the connecting magnetic flux. Besides the plane $z=0$, the
equipotential $\mathcal V=0$ contains the surface $R_2=\lambda
R_1$. For $\lambda\ne1$, this surface is the sphere
\be
(x-c_{\rm n})^2+y^2+z^2=r_{\rm n}^2,
\ee
where 
\be c_{\rm n}=\frac{a}{1-\lambda^2},
\quad
r_{\rm n}=\frac{a\lambda}{|1-\lambda^2|}.
\ee
%Here $c_{\rm n}$ is the sphere's centre coordinate and $r_{\rm n}$ its radius.
To evaluate the flux, take $0<\lambda<1$, so that the sphere
encloses dipole 2 and excludes dipole 1. The field is normal to this
equipotential. The scalar potential drops along a field line; after
crossing the equipotential the potential decreases below zero, so the
line cannot return to the sphere or reach infinity, where the
potential is again zero. See Fig.~1 (left panel) in the main text.  We can
calculate the connecting flux by integrating over the hemisphere
$S_+$ with $z>0$, where the field lines point outward the
sphere. The flux crossing $S_+$ therefore connects from dipole 2 to
dipole 1.

Since $S_+$ avoids both dipole singularities, its flux can be
evaluated by Stokes' theorem. Introducing the vector potential
$\mathbf A=\sum_{i=1}^{2}\boldsymbol\mu_i\times\mathbf R_i/R_i^3$, for
which $\mathbf B=\boldsymbol\nabla\times\mathbf A$, we obtain
\be
\Phi_L=\int_{S_+}\mathbf B\cdot\mathrm d\mathbf S
=\oint_C\mathbf A\cdot\mathrm d\boldsymbol\ell.
\ee
We integrate the vector potential along the intersection between the
equipotential sphere and $xy$ plane. Parameterize the boundary by
$(x,y,z)=(c_{\rm n}+r_{\rm n}\cos\psi,r_{\rm n}\sin\psi,0)$, where
$0\le\psi<2\pi$ increases counterclockwise as viewed from $+z$,
consistently with the outward surface normal. On $C$, the relation
$R_2=\lambda R_1$ reduces the vector potential to $\mathbf
A|_C=-\mu_1a\hat{\mathbf y}/R_1^3$. The flux consequently becomes the
one-dimensional integral
\be
\Phi_L=-\frac{\mu_1}{a}\lambda(1-\lambda^2)^2
\int_0^{2\pi}
\frac{\cos\psi\,\mathrm d\psi}
{(1+\lambda^2+2\lambda\cos\psi)^{3/2}}.
\ee
Evaluating this integral gives the following expression:
\be
\Phi=\frac{2\mu_1(1+\lambda)}{a}
\left[(1+\lambda^2)E(m)-(1-\lambda)^2K(m)\right],
\ee
where
\be
m=\frac{4\lambda}{(1+\lambda)^2}.
\ee
Here $K$ and $E$ are the complete elliptic integrals of the first and
second kinds, respectively, with parameter $m$:
\ba
&&K(m)=\int_0^{\pi/2}\frac{\mathrm d\chi}{\sqrt{1-m\sin^2\chi}},\\
&&E(m)=\int_0^{\pi/2}\sqrt{1-m\sin^2\chi}\,\mathrm d\chi.
\ea
The dimensionless flux factor is then
%Substitution into the adopted normalization gives, for every $\lambda>0$ with $\lambda\ne1$,
\be
f_\Phi(\lambda)=\frac{2(1+\lambda)^3}{\pi\lambda^2}
\left[(1+\lambda^2)E(m)-(1-\lambda)^2K(m)\right].
\ee
This is the expression used for plotting Fig.~2.

For $\lambda=1$ (with $\mu_1=\mu_2\equiv\mu$), the condition $R_2=R_1$
now defines the midplane $x=a/2$, whose upper half-plane $z>0$ is the
corresponding flux cross-section. The field there points from dipole 2
towards dipole 1, so the positive linked flux follows directly from
\be
\mathbf B=-\frac{3\mu a z}{(a^2/4+y^2+z^2)^{5/2}}\hat{\mathbf x},
\ee
and
\be
\Phi_L=3\mu a\int_{-\infty}^{\infty}\!\mathrm dy
\int_0^\infty\!\frac{z\,\mathrm dz}{(a^2/4+y^2+z^2)^{5/2}}
=\frac{8\mu}{a}.
\ee
Since $r_{m2}=a/2$, we find
\be
f_\Phi(1)=\frac{8\mu/a}{\pi(a/2)^2\mu/a^3}
=\frac{32}{\pi}.
\ee

The equal-moment value is the continuous limit of the elliptic
expression; it supplies the value at $m=1$, where $K(m)$ itself
diverges. Source interchange gives $f_\Phi(\lambda)=\lambda
f_\Phi(1/\lambda)$, allowing values above unity to be obtained from
those below unity. In the weak-moment limit ($\lambda\ll 1$),
we have
\be
f_\Phi=3(1+\lambda)^2\left[1-{\lambda^2\over 8}+{\cal O}(\lambda^4)\right],
\quad (\lambda\ll 1).
%\to3$ as $\lambda\to0$.
\ee


\begin{thebibliography}{99}

\bibitem{Go69}
P.~Goldreich and D.~Lynden-Bell,
Astrophys.~J. \textbf{156}, 59 (1969).

\bibitem{Cl96}
J.~T. Clarke \textit{et al.},
Science \textbf{274}, 404 (1996).

\bibitem{Lanza09}
A.~F. Lanza,
Astron.~Astrophys. \textbf{493}, 1127 (2009).

\bibitem{Laine12}
R.~O. Laine and D.~N.~C. Lin,
Astrophys.~J. \textbf{745}, 2 (2012).

\bibitem{Strugarek14}
A.~Strugarek, A.~S. Brun, S.~P. Matt, and V.~R\'eville,
Astrophys.~J. \textbf{795}, 86 (2014).

\bibitem{Strugarek16}
A.~Strugarek,
Astrophys.~J. \textbf{833}, 140 (2016).

\bibitem{Strugarek17}
A.~Strugarek, A.~S. Brun, S.~P. Matt, and V.~R\'eville,
Astrophys.~J. \textbf{847}, 36 (2017).

\bibitem{Hansen01}
B.~M.~S. Hansen and M.~Lyutikov,
Mon.~Not.~R.~Astron.~Soc. \textbf{322}, 695 (2001).

\bibitem{Lyutikov19}
M.~Lyutikov,
Mon.~Not.~R.~Astron.~Soc. \textbf{483}, 4168 (2019).

\bibitem{Most20}
E.~R. Most and A.~A. Philippov,
Astrophys.~J. \textbf{901}, L26 (2020).

\bibitem{Most22}
E.~R. Most and A.~A. Philippov,
Mon.~Not.~R.~Astron.~Soc. \textbf{515}, 2710 (2022).

\bibitem{Most23}
E.~R. Most and A.~A. Philippov,
Astrophys.~J. \textbf{955}, 24 (2023).

\bibitem{Mahlmann25}
J.~F. Mahlmann and A.~M. Beloborodov,
Astrophys.~J. \textbf{981}, L17 (2025).

\bibitem{Beloborodov21}
A.~M. Beloborodov,
Astrophys.~J. \textbf{922}, L7 (2021).

\bibitem{Lai12}
D.~Lai,
Astrophys.~J. \textbf{757}, L20 (2012).

\bibitem{Cherkis21}
S.~A. Cherkis and M.~Lyutikov,
Astrophys.~J. \textbf{910}, 107 (2021).

\bibitem{aly85}
J.~J. Aly,
Astron.~Astrophys. \textbf{143}, 19 (1985).

\bibitem{aly90}
J.~J. Aly and J.~Kuijpers,
Astron.~Astrophys. \textbf{227}, 473 (1990).

\bibitem{van94}
A.~A. van Ballegooijen,
Space Sci.~Rev. \textbf{68}, 299 (1994).

\bibitem{Lo95}
R.~V.~E. Lovelace, M.~M. Romanova, and G.~S. Bisnovatyi-Kogan,
Mon.~Not.~R.~Astron.~Soc. \textbf{275}, 244 (1995).

\bibitem{Uz02}
D.~Uzdensky, A.~K\"onigl, and C.~Litwin,
Astrophys.~J. \textbf{565}, 1191 (2002).

\bibitem{LeeChiang17}
E.~J. Lee and E.~Chiang,
Astrophys.~J. \textbf{842}, 41 (2017).

\bibitem{PuLai19}
B.~Pu and D.~Lai,
Mon.~Not.~R.~Astron.~Soc. \textbf{488}, 3566 (2019).

\bibitem{Drell65}
S.~D. Drell, H.~M. Foley, and M.~A. Ruderman,
J.~Geophys.~Res. \textbf{70}, 3131 (1965).

\bibitem{CorsoLai26}
N.~Corso and D.~Lai,
Astrophys.~J. \textbf{1002}, 1 (2026).

\bibitem{CorsoLai26b}
N.~Corso and D.~Lai,
Astrophys.~J. submitted (2026); arXiv:2610.30506

\bibitem{Mae12}
H.~Maehara \textit{et al.},
Nature \textbf{485}, 478 (2012).

\bibitem{Sh08}
E.~Shkolnik, D.~A. Bohlender, G.~A.~H. Walker, and A.~Collier Cameron,
Astrophys.~J. \textbf{676}, 628 (2008).

\bibitem{Pi11}
I.~Pillitteri \textit{et al.},
Astrophys.~J. \textbf{741}, L18 (2011).

%\bibitem{fa12}
%R.~Fares \textit{et al.}, Mon.~Not.~R.~Astron.~Soc. \textbf{423}, 1006 (2012).

%\bibitem{Mi12}
%B.~P. Miller, E.~Gallo, J.~T. Wright, and A.~K. Dupree,
%Astrophys.~J. \textbf{747}, 135 (2012).

\bibitem{Callingham24}
J.~R. Callingham, B.~J.~S. Pope, R.~D. Kavanagh, S.~Bellotti, S.~Daley-Yates, et al.,
Nature Astronomy \textbf{8}, 1359 (2024).

\bibitem{Ilin25}
E.~Ilin, H.~K. Vedantham, K.~Poppenh\"ager, S.~Bloot, J.~R. Callingham, A.~Brandeker, and H.~Chakraborty,
Nature \textbf{643}, 645 (2025).

\bibitem{Pineda26}
J.~S. Pineda, S.~Bellotti, J.~Villadsen, A.~Vidotto, J.~Morin, and C.~P. Folsom,
Astrophys.~J. \textbf{997}, 172 (2026).

\end{thebibliography}
\end{document}